\documentclass[prl,showpacs,preprint]{revtex4}%
\usepackage{amsmath}
\usepackage{graphicx}
\usepackage{amsfonts}
\usepackage{amssymb}%
\begin{document}
\title{Possible approach to improve sensitivity of a Michelson interferometer}
\author{Jian Fu}
\affiliation{State Key Lab of Modern Optical Instrumentation, Department of Optical
Engineering, Zhejiang University, Hangzhou 310027, China}

\begin{abstract}
We propose a possible approach to achieve an $1/N$ sensitivity of Michelson
interferometer by using a properly designed random phase modulation. Different
from other approaches, the sensitivity improvement does not depend on
increasing optical powers or utilizing the quantum properties of light.
Moreover the requirements for optical losses and the quantum efficiencies of
photodetection systems might be lower than the quantum approaches and the
sensitivity improvement is frequency independent in all detection band.

\end{abstract}

\pacs{04.80.Nn, 42.87.Bg, 07.60.Ly}
\maketitle

Michelson interferometer plays an important role in ultimate sensitivity
measurements. Especially, large-scale interferometric gravitational wave
detectors \cite{Thorne} such as the Laser Interferometric Gravity-wave
Observatory (LIGO) is a Michelson interferometer with arm cavities and power
recycling \cite{Barish}. In the measurements, the ultimate sensitivity is
conventionally bounded by the quantum nature of the electromagnetic field. It
has been shown that the so-called shot noise is due to the vacuum fluctuations
coupled to the interferometer and the radiation pressure noise is due to the
random motion of the mirrors induced by the radiation pressure fluctuations
\cite{Caves}. However, these standard quantum limits (SQL) are not as
fundamental as the Heisenberg limits, and can be beaten by using quantum
entanglement \cite{Giovannetti} and squeezing \cite{Xiao}. Benefiting from a
nonlocal correlation, quantum entanglement can achieve an $1/N$ sensitivity
with a $\sqrt{N}$ precision improvement ($N$ being the number of photons
employed) over the classical strategies \cite{Giovannetti}. But there is an
enormous difficulty in the quantum-enhanced measurement, that is usually very
complicated to realize multi-particle quantum entanglement even as few as 5 or
6 particles \cite{Zhao}. Other quantum approaches using squeezed states
\cite{Xiao,Bondurant} or Fock states \cite{Holland} can also achieve the $1/N$
sensitivity \cite{Giovannetti}. However, these approaches for sensitivity
enhancement beyond the SQL require optical losses to be very low and the
quantum efficiencies of photodetection systems to be very high \cite{Buonanno}.

Recently, \textquotedblleft mode-entangled states\textquotedblright\ based on
the transverse modes of classical optical fields propagating in multimode
waveguides are proposed as classical simulations of quantum entangled states
\cite{Fu}. It is interesting that the mode-entangled states can also exhibit
the nonlocal correlation, such as the violation of Bell's inequality. By using
the classical simulation of $N$-particle quantum entangled states, an
interferometer that can beat the standard quantum limit was proposed in Ref.
\cite{Fu1}. Similar to the quantum-enhanced measurements, the interferometer
can also achieve an $1/N$ sensitivity with a $\sqrt{N}$ precision improvement
over ordinary interferometers. It is noticeable that the nonlocal correlation
similar to quantum entanglement might be caused by a random phase mechanism
\cite{Fu2}. It might also be a possible mechanism for improving the
sensitivity and precision of interferometers \cite{Fu1}.

In this paper, we propose a possible approach to improve the sensitivity of
Michelson interferometer by using a random phase modulation. In the scheme, we
will divide each sampling period into $N$ time slots. The measurement in each
time slot can be regarded as an independent measurement. By modulating
properly designed random phase sequences in time slots, the $N$ measurements
will become similar to the measurements employed $N$-field mode-entangled
states \cite{Fu1}. At last, by performing a correlation analysis on these
measurement results, it is in principle possible to achieve an $1/N$ sensitivity.

Before turning to a detailed analysis of the scheme, it is useful to discuss
an ordinary Michelson interferometer. A classical coherent beam with $N_{p}$
average photons enters impinges on a semi-transparent mirror (i.e. a beam
splitter), which divides it into a reflected and a transmitted beam. These two
beams travel along different paths and then are reflected back by two mirrors.
At last they are recombined by the same beam splitter. By measuring the
intensity difference of the two output beams, one can recover the phase
$\theta=4\pi\Delta L/\lambda$ ( $\Delta L$ being the path difference) between
the two optical paths with a statistical error proportional to $1/\sqrt{N_{p}%
}$. This is the SQL due to the quantized nature of electromagnetic field and
the Poissonian statistics of classical light.

Here, a Michelson interferometer with a minor change to achieve an $1/N$
sensitivity is proposed, and the scheme is shown in Fig. 1. In the scheme, a
phase modulator $\varphi$ is mounted on one of the optical paths. This is only
a different between the scheme and an ordinary Michelson interferometer.
Suppose that a time dependent extremely small phase $\theta=4\pi\Delta
L/\lambda$ has an upper frequency limit $f_{u}$. According to Nyquist sampling
theory, we can determine a sampling period $T_{s}=1/2f_{u}$ to measure the
phase difference. Then we divide each sampling period into $N$ time slots. The
measurement in each time slot can be regarded as an independent measurement.
Further, we divide each time slot into $M$ phase units. In each phase unit, a
random phase $\varphi$ uniformly distributed in $\left[  0,2\pi\right]  $ is
modulated by the phase modulator. Thus, we obtain $N$ random phase sequences
of length $M$ that can be written as $\left\{  \varphi_{j}^{i},\left(
i=1...N,j=1...M\right)  \right\}  $. By measuring the intensity difference $I$
of the two output beams in each phase unit, we can obtain
\begin{equation}
I_{j}^{i}=\frac{N_{p}}{MN}\cos\left(  \theta+\varphi_{j}^{i}\right)  .
\label{e1}%
\end{equation}
Due to the random phase $\varphi_{j}^{i}$ uniformly distributed in $\left[
0,2\pi\right]  $, the intensity difference $I_{j}^{i}$ is also randomly varied
in the range of $\pm N_{p}/MN$. Thus, we can obtain $N$ random intensity
difference sequences $\left\{  I_{j}^{i},\left(  i=1\ldots N,j=1\ldots
M\right)  \right\}  $. The relation of the time slots, the phase and intensity
difference sequences is shown in Fig. 2.

In order to extract the information of $\theta$ from the random intensity
difference sequences $\left\{  I_{j}^{i}\right\}  $, we properly design the
random phase sequence of the $N$th time slot as
\begin{equation}
\varphi_{j}^{N}=2\pi-\left(  \sum_{i=1}^{N-1}\varphi_{j}^{i}\right)
\operatorname{mod}2\pi.\label{e2}%
\end{equation}
Then a correlation analysis is performed on the intensity difference
sequences, and the correlation function can be written as
\begin{align}
S_{N}\left(  \theta\right)   &  =\frac{1}{M}\sum_{j=1}^{M}\left[
{\displaystyle\prod\limits_{i=1}^{N}}
I_{j}^{i}\right]  \label{e3}\\
&  =\frac{1}{M}\left(  \frac{N_{p}}{MN}\right)  ^{N}\sum_{j=1}^{M}\left[
{\displaystyle\prod\limits_{i=1}^{N}}
\cos\left(  \theta+\varphi_{j}^{i}\right)  \right]  \nonumber\\
&  =\frac{2}{M}\left(  \frac{N_{p}}{2MN}\right)  ^{N}\sum_{j=1}^{M}\left\{
\cos\left(  N\theta+\sum_{i=1}^{N}\varphi_{j}^{i}\right)  +\cos\left[  \left(
N-2\right)  \theta+\sum_{i=1}^{N}\varphi_{j}^{i}-2\varphi_{j}^{1}\right]
+\ldots\right\}  .\nonumber
\end{align}
By using $\sum_{i=1}^{N}\varphi_{j}^{i}=2\pi$ obtained from Eq. (\ref{e2}), we
can obtain
\begin{equation}
S_{N}\left(  \theta\right)  =2\left(  \frac{N_{p}}{2MN}\right)  ^{N}\left\{
\cos\left(  N\theta\right)  +\frac{1}{M}\sum_{j=1}^{M}\left[  \cos\left(
N\theta-2\theta-2\varphi_{j}^{1}\right)  +\ldots\right]  \right\}  ,\label{e4}%
\end{equation}
where the expression of $\sum_{j=1}^{M}\ldots$ contains $2^{N-1}-1$ terms that
are all random-phase cosine functions. If the random phases are uniformly
distributed in $\left[  0,2\pi\right]  $, the sum of the random-phase cosine
terms equals zero.

By using orthogonal pseudo-random number (PN) sequence techniques, we can
completely eliminate the terms of $\sum_{j=1}^{M}\ldots$. In all PN sequences,
maximal-length linear feedback shift-register sequence (M-sequence) is often
used as spread-spectrum sequence and can simultaneously exhibit good weight
distribution and correlation properties. Here, we use the M-sequence technique
to generate the random phase sequences. After given a generating polynomial of
order $n$, $2^{n}-1$ different M-sequences of length $2^{n}-1$ can be obtained
by shifting bits. By properly chosen from these sequences, the $N-1$ random
sequences of length $M=2^{n}$ (a code 0 or 1 added to balance their numbers)
can be obtained. Then, by alternately mapping the codes $0$ to the phases $0$
and $\pi$, and $1$ to $\pi/2$ and $3\pi/2$, we obtain the $N-1$ random phase
sequences in which the phases are assigned with four discrete values
$0,\pi/2,\pi$ and $2\pi/3$. By using Eq. (\ref{e2}), the random phase sequence
of the $N$th time slot can be obtained. By using the $N$ random phase
sequences, we eliminate the terms of $\sum_{j=1}^{M}\ldots$ and obtain the
correlation function
\begin{equation}
S_{N}\left(  \theta\right)  =2\left(  \frac{N_{p}}{2MN}\right)  ^{N}%
\cos\left(  N\theta\right)  . \label{e5}%
\end{equation}
This result shows a sensitivity of the order $1/N$ for the measurements of
small phase $\theta$.

In the quantum-enhanced measurement, a $\sqrt{N}$ precision improvement over
the classical strategies can be achieved, with the concomitant improvement in
sensitivity. However, in the scheme, the $\sqrt{N}$ precision improvement
might not be achieved due to the complication of the random phase modulation.
Its precision is mainly limited by two unavoidable sources of errors. The
first source is the shot noise. In each phase unit, the phase error is
proportional to $\sqrt{MN/N_{p}}$ due to $N_{p}/MN$ average photons. In the
correlation analysis, the variance associated with $\frac{1}{M}\sum_{j=1}%
^{M}\ldots$ is given by $\sqrt{N/N_{p}}$. By using the correlation function,
the phase error $\Delta\theta$ can be obtained from error propagation,
$\Delta\theta=\Delta S_{N}/\left\vert \partial S_{N}/\partial\theta\right\vert
$, it is easy to see that it scales as $1/N$. Therefore we can obtain the shot
noise limit as $\Delta\theta=1/\sqrt{NN_{p}}$. The second source is a phase
error induced by the phase modulation. When a light beam is modulated by an
electro-optical phase modulator such as LiNbO$_{3}$ modulator, the phase error
$\Delta\varphi$ is induced by the voltage fluctuations of control
signals.\smallskip\ The phase error will lead to the intensity fluctuations
due to the terms of $\sum_{j=1}^{M}\ldots$ and the phase sum error $\sum
_{i=1}^{N}\varphi_{j}^{i}=2\pi\pm N\Delta\varphi$. The phase precision limit
can be estimated to be $(\sqrt{2^{N-1}\left(  N-1\right)  /MN}+1)\Delta
\varphi$. At last, we can obtain the overall phase error
\begin{equation}
\Delta\theta=\frac{1}{\sqrt{NN_{p}}}+\left(  \sqrt{\frac{2^{N-1}\left(
N-1\right)  }{MN}}+1\right)  \Delta\varphi. \label{e6}%
\end{equation}
Although, a $\sqrt{N}$ precision improvement of the shot noise limit is
achieved in the scheme, a new phase error is induced by the phase modulation.

In order to reduce the new phase error $\Delta\varphi$, we propose a
configuration of optical phase-locked loop (OPLL) as shown in Fig. 3. The OPLL
technique is generally applied to control the phase error of two different
lasers in optical PSK homodyne transmission systems and easily realized
without any rigorous requirements for optical losses and photodetection
systems \cite{Kazovsky,Norimatsu,Malyon}. In this scheme, two optical beams
are split from the fore-and-aft places of the phase modulator and combined by
a semi-transparent mirror. After measured the intensity difference, the signal
is further processed by a gain adjustable loop amplifier and a loop filter. To
complete the loop, the loop filter is connected to a control voltage generator
that generates suitable voltages to control the phase modulator. The
performance of the OPLL depends on the properties of the input optical beams
and loop design. In this scheme, the shot noise might be the main source of
the phase error. When the loop is locked, the phase error contributed by the
shot noise can be written as $\Delta\varphi=\sqrt{N_{0}B_{L}}/A$, where
$N_{0}$ is the power spectral density of the shot noise, $B_{L}$ is the one
sided loop bandwidth, and $A=2rRP$ is the gain of the intensity difference
detectors with the receiver transimpedance $r$, the photodetector responsivity
$R$ and the received optical powers $P$. The phase locking action can
dramatically suppress the phase error for a narrow loop bandwidth and a high
value of $A^{2}/N_{0}$ \cite{Win}. As reported in Ref. \cite{Kazovsky}, the
phase error of two different lasers might be controlled below $0.3%
{{}^\circ}%
$ for signal powers $P_{s}\geqslant-62dBm$. In this scheme, the phase error
limit should be less than that for without the influence of frequency
perturbation. Further refinement of the OPLL, such as optimal loop bandwidth,
might allow to improve the phase precision until nearly reaching the $\sqrt
{N}$ precision improvement over ordinary interferometers.

In many measurements with optical interferometers, the ultimate sensitivity is
required. By using this scheme, we can achieve an $1/N$ sensitivity with few
changes on the optical interferometers. Consider the detection band of $40Hz$
to $5kHz$, the sampling period $T_{s}=0.1ms$ can be determined. If an $1/10$
sensitivity is required, each sampling period is divided into $10$ time slots.
After given the $8$th-order generating polynomial $f\left(  x\right)
=1+x^{2}+x^{3}+x^{4}+x^{8}$, $255$ M-sequences of length $255$ are obtained.
By properly chosen and calculated, $10$ random phase sequences of length
$M=256$ are obtained to eliminate the terms of $\sum_{j=1}^{M}\ldots$. Then,
we can obtain that the rate requirement for the phase modulator is $25.6MHz$.
It is very easy to reach the requirement for LiNbO$_{3}$ modulator that is a
fairly mature commercial technology. Further, we can estimate the phase error
$\Delta\theta\sim0.7%
{{}^\circ}%
$ induced by the phase modulation (assumed the phase error as Ref. [13]).
After careful study, we find the generating polynomial order $n\geqslant N-2$.
This leads to a difficulty to further improve the sensitivity, that is the
number of phase units require exponentially increasing with the sensitivity
improvement. It might be reduced by using some new PN sequences.

In this paper, we have discussed a possible approach to improve the
sensitivity for a Michelson interferometer. Different from other approaches,
the sensitivity improvement does not depend on increasing optical powers or
utilizing the quantum properties of light. Moreover its requirements for the
optical losses and the quantum efficiencies of photodetection systems might be
lower than the quantum approaches and the sensitivity improvement is frequency
independent in all detection band. The approach might be applicable to some
laser interferometers for gravity-wave detection. Most gravity-wave laser
interferometers, such as LIGO, rely on sophisticated power and signal
recycling schemes with carrier/sideband measurement signatures. Therefore, a
readout and control scheme compatible with both the correlation analysis and
the recycle of power and signal deserves further study. Besides this
application, the approach can be generally applied to various types of
interferometers without any rigorous requirements of optical powers or the
quantum properties of light.

This work was supported by the National Natural Science Foundation of China
under Grant No. 60407003.

\begin{description}
\item \pagebreak

\item[Fig. 1:] The scheme of the Michelson interferometer to achieve higher sensitivity.

\item[Fig. 2:] The relation of the time slots, the phase and intensity
difference sequences.

\item[Fig. 3:] The configuration of optical phase-locked loop to control the
phase error, LF: Loop Filter, CVG: Control Voltage Generator, PNG:
Pseudo-random Number Generator, PM: Phase Modulator.
\end{description}


\bigskip

\begin{thebibliography}{99}                                                                                               %


\bibitem {Thorne}K. Thorne, Rev. Mod. Phys. \textbf{52}, 285 (1980); L. Ju, D.
G. Blair, and C. Zhao, Rep. Prog. Phys. \textbf{63}, 1317 (2000).

\bibitem {Barish}B. Barish and R. Weiss, Phys. Today \textbf{52}, 44 (1999);
A. Abramovici, and et al., Science \textbf{256}, 325 (1992).

\bibitem {Caves}C. M. Caves, Phys. Rev. Lett. \textbf{45}, 75 (1980); V. B.
Braginsky, and Y. I. Vorontsov, Sov. Phys. Usp. \textbf{17}, 644 (1975); W. A.
Edelstein, J. Hough, J. R. Pugh, J. Martin, J. Phys. E \textbf{11}, 710 (1978).

\bibitem {Giovannetti}V. Giovannetti, S. Lloyd, L. Maccone, Science
\textbf{306}, 1330 (2004).

\bibitem {Xiao}M. Xiao, Ling-An Wu, and H. J. Kimble, Phys. Rev. Lett.
\textbf{59}, 278 (1987).

\bibitem {Zhao}Z. Zhao, and et al., Nature \textbf{430,} 55 (2004).

\bibitem {Bondurant}R. S. Bondurant and J. H. Shapiro, Phys. Rev. D
\textbf{30}, 2548 (1984).

\bibitem {Holland}M. J. Holland and K. Burnett, Phys. Rev. Lett. \textbf{71},
1355 (1993); J. Jacobson, and et al., Phys. Rev. Lett. \textbf{74}, 4835 (1995).

\bibitem {Buonanno}A. Buonanno and Y. Chen, Phys. Rev. D \textbf{69}, 102004
(2004); A. Buonanno and Y. Chen, Phys. Rev. D \textbf{64} 042006 (2001); T.
Kim, Y. Ha, and et al., Phys. Rev. A \textbf{60}, 708 (1999).

\bibitem {Fu}J. Fu, Z. J. Si, S. F. Tang, and et al., Phys. Rev. A
\textbf{70}, 042313 (2004).

\bibitem {Fu1}J. Fu, quant-ph/0604103.

\bibitem {Fu2}J. Fu and X. Zhang, quant-ph/0506125.

\bibitem {Kazovsky}L. G. Kazovsky, J. Lightwave Technol. \textbf{8}, 1414
(1990); D. A. Atlas and L. G. Kazovsky, IEEE Photon. Technol. Lett.
\textbf{2}, 367 (1990).

\bibitem {Norimatsu}S. Norimatsu, K. Iwashita, and K. Noguchi, Electron. Lett.
\textbf{26}, 648 (1990); S. Norimatsu and K. Iwashita, J. Lightwave Technol.
\textbf{9}, 1367 (1991); B. Wandernoth, Electron. Lett. \textbf{28}, 387
(1992); J. M. Kahn, IEEE Photon. Technol. Lett. \textbf{1}, 340 (1989).

\bibitem {Malyon}D. J. Malyon, Electron. Lett. \textbf{20}, 281 (1984); L. G.
Kazovsky, J. Lightwave Technol. \textbf{LT-4}, 182 (1986); D. J. Malyon, D. W.
Smith, and R. Wyatt, Electron. Lett. \textbf{22}, 421 (1986); J. M. Kahn, and
et al., IEEE Photon. Technol. Lett. \textbf{2}, 285 (1990).

\bibitem {Win}M. Z. Win, C-C. Chen, and R. A. Scholtz, Proceedings of SPIE
\textbf{1417}, 40 (1991).
\end{thebibliography}
\end{document}